\documentclass[multphys,vecphys]{svmult}

\usepackage{makeidx}         
\usepackage{graphicx}        
\usepackage{multicol}        
\usepackage[bottom]{footmisc}

\makeindex             

\begin{document}

\title*{Globular Clusters in Dwarf Galaxies}
\author{Bryan W. Miller}
\institute{Gemini Observatory, Casilla 603, La Serena, Chile\\
\texttt{bmiller@gemini.edu}}
%
%
\maketitle

\begin{abstract}
Recent work on globular cluster systems in dwarf galaxies outside the
Local Group is reviewed.  Recent large imaging surveys with the {\it
  Hubble Space Telescope} and follow-up spectroscopy with 8-m class
telescopes now allow us to compare the properties of massive star
clusters in a wide range of galaxy types and environments.  This body
of work provides important constraints for theories of galaxy and star
cluster formation and evolution.
\end{abstract}

\section{Introduction}
\label{sec:intro}

Studies of globular clusters (GCs) in dwarf galaxies provide very
important insights into galaxy formation, the formation and evolution
of GCs, and the relationship between GCs and nuclei. Comparisons of
the properties of star clusters in different types of galaxies can
test the theories of galaxy formation.  In hierarchical scenarios of
galaxy formation dwarf-size galaxies form first and then merge into
larger systems.  If star cluster formation coincided with galaxy
formation, then a significant fraction of the star clusters in massive
galaxies should have been formed in dwarfs.  In this case the star
clusters in dwarf galaxies in dense environments should be at least as
old and metal-poor as the oldest star clusters in giant galaxies.
However, recently evidence has mounted that stellar populations in
surviving low mass galaxies are younger than in giant
ellipticals\cite{treu05}.  In this ``downsizing'' view the dwarf
galaxies formed after the giants or at least had their star formation
rates suppressed at early times. A signature of downsizing would be
that the star clusters in dwarfs are younger than those in giant
galaxies.

Another question that star clusters can help answer is the
relationship between dwarf irregular (dI) and dwarf elliptical (dE)
galaxies.  All dwarf galaxies must have formed with substantial gas
fractions like today's dI galaxies.  However, in massive local
galaxies clusters the majority of the dwarfs are gas-free,
smooth-isophote dEs.  The differences may be due to environment or dIs
may get transformed into dEs by gas stripping, supernovae winds, or
galaxy interactions.  A comparison of the star clusters in the two
types of dwarfs provides insight into the processes that shaped these
galaxies and into why some dEs form nuclei.

In addition, the shape of the initial mass function of star clusters
and how it evolves is not well understood.  There are still debates
about whether the form of initial mass function is a single or broken
power-law (resulting in a log-normal distribution in magnitudes) and
about the effects of various destruction processes
\cite{baum98}\cite{fz01}\cite{vesp01}.  By comparing the present-day
mass functions in dwarf galaxies with those in giant galaxies it may
be to disentangle the destructive processes and therefore determine
the shape of the initial star cluster mass function.

This paper reviews the properties of star cluster systems in dwarf
galaxies outside of the Local Group.  Large imaging surveys with the
{\it Hubble Space Telescope} are now starting to provide us with
statistically significant samples of GCs in dEs and dIs in different
environments.  Follow-up spectroscopy with 8-m class telescopes are
now providing complementary results on the ages, metallicities,
abundance ratios, and kinematics of GCs and nuclei in dwarf galaxies.

\section{Radial Distributions}
\label{sec:rad}

\begin{figure}
\centering
\includegraphics[width=6.5cm,angle=90]{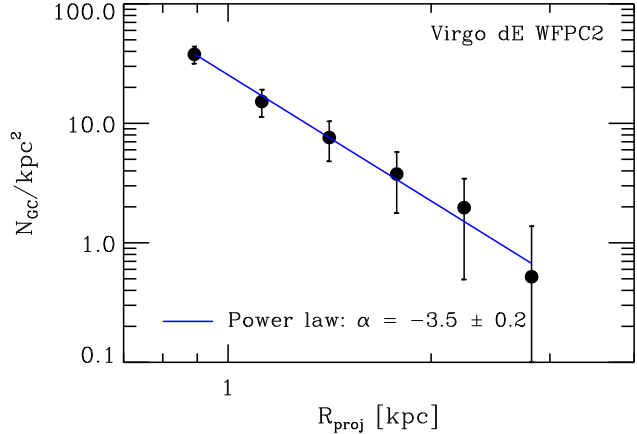}
%
%
\caption{The projected radial distribution of globular clusters in the
Virgo Cluster sample of the WFPC2 dE Snapshot Survey.  The data is
well-fit by a single power-law with a slope $\alpha = -3.5 \pm
0.2$.}
\label{fig:rad}       
\end{figure}

A common problem when studying the globular cluster systems (GCSs) of
dwarf galaxies is that any given galaxy generally has too few clusters
to draw broad conclusions.  Therefore, the standard approach is to
combine the clusters from a large number of galaxies into a ``master''
dE GCS.  Various studies have found that the radial distribution of
GCs in dEs follows that of the background light and that it has a
power-law form with a slope ranging between $-1.6$ and $-3.5$
\cite{harris86}\cite{durrell96}\cite{minniti96}\cite{sharina05}.
Figure~\ref{fig:rad} show the background-subtracted radial
distribution of GCs from the WFPC2 dE Snapshot Survey.  The
distribution is a power-law with $\alpha=-3.5\pm0.2$.

An alternative way of characterizing the radial distribution is to
scale the projected radius of each cluster by the scale length of the
host galaxy. This allows a direct comparison between the GC and
background light distributions for a combined sample with galaxies of
varying sizes.  Data from the WFPC2 dE Snapshot Survey have shown that
the radial distribution of the complete sample of GCs follows the
background light extremely well \cite{lotz01}.  However, the
distribution of the GCs with $M_V < -8$ shows a deficit at small radii
that may be the result of dynamical friction.  The dynamical friction
timescales in dEs are short enough that the merging of GCs via this
process is one avenue of producing nuclei
\cite{hg98}\cite{ohlin00}\cite{lotz01}\cite{bekki06}.  However, simple
dynamical friction calculations over-predict the luminosities of the
nuclei so other processes may be counteracting it \cite{lotz01}.

\section{Luminosity Functions}
\label{sec:gclf}

The observed GC luminosity function (GCLF) gives the present-day GC
mass function if the ages and metallicities ($M/L$ ratios) of the
clusters are known.  Modeling the processes that can destroy GCs in
dwarfs, mainly two-body relaxation with stellar evolution will
hopefully allow us to determine the initial GCLF.   

The GCLF in dEs has been measured recently for galaxies in nearby
groups and the field \cite{sharina05} and in the Virgo and Fornax
Clusters \cite{ml06}\cite{jordan06}.  In the Virgo Cluster the
combined GCLF from WFPC2 data plotted as a function of magnitude is
fit by a $t_5$ distribution with a peak at $M_V^0 = -7.3 \pm 0.1$.
This is consistent with a GCLF peak of $M_V^0 \approx -7.5\pm 0.3$ in
VCC~1087 \cite{beasley06} and in the nearby group sample there is a
peak at $M_V^0=-7.4$ but after a small decline the numbers continue to
rise at fainter magnitudes \cite{sharina05}.

A key issue is whether the GCLF peak in dEs is the same as the peak
seen in old. metal-poor GCs in giant galaxies.  Di~Criscienzo et
al. have recently compared the GCLFs for the Milky Way, M31, and
several giant ellipticals in Virgo using consistent selection criteria
and distance scale \cite{dicris06}.  Fits to a $t_5$ distribution give
very consistent peaks with an average value of $M_V^0 = -7.66\pm0.2$.
The GCLF peak for the dEs is consistent with this value, suggesting
that the GCLF peak for old, metal-poor GC populations is nearly
universal.  However, there is a suggestion that the peak in dEs is
$\sim0.3$ mag fainter than in giant galaxies, perhaps as result of less
efficient disk shocking\cite{mil06}.

The GCLF can also be plotted as a function of luminosity rather than
magnitude.  In this representation the peaks discussed above
correspond to breaks in a power-law distribution.  The bright-end
GCLF in dwarfs is consistent with $\phi(L)/L \propto L^{\alpha}$ with
$\alpha \sim -1.9$, similar to the slopes of the mass functions of
Galactic molecular clouds and the luminosity functions of very young
star clusters in starburst galaxies\cite{mil06}.

\begin{figure}
\centering
\includegraphics[width=6.5cm,angle=90]{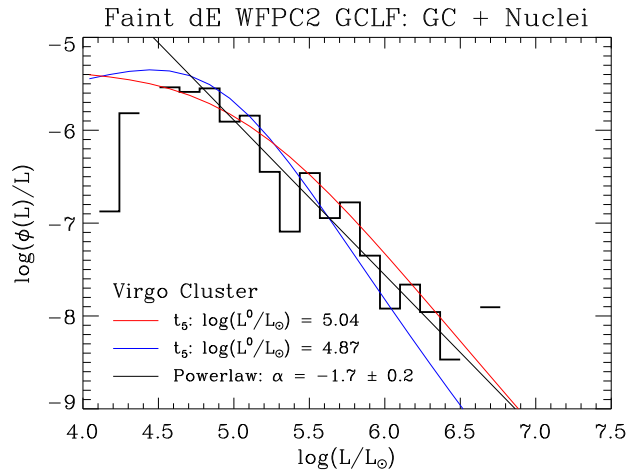}
\caption{Background-subtracted luminosity function for GCs and nuclei
  for Virgo galaxies in the WFPC2 Virgo sample with $M_V >
  -15.75$. For $\log(L/L_{\odot}) > 4.8$ the data is well-fit by a
  power-law with $\alpha = -1.7\pm0.2$.  The data is also reasonably
  consistent with the $t_5$ fit to the entire WFPC2 sample which has a
  peak at $\log(L/L_{\odot}) = 4.87$ (blue line).  The red line is the
  best fitting $t_5$ function to the faint-galaxy GCLF and it has a
  brighter and broader peak than the standard GCLF.}
\label{fig:gclffaint}       
\end{figure}

Recently, van den Bergh has proposed that the GCLF for galaxies with
$M_V > -16$ is a single power-law, without a break\cite{vdb06}.  The
WFPC2 dE Snapshot data is also consistent with this but low number
statistics make it difficult to distinguish various models
(Figure~\ref{fig:gclffaint}).



\section{Colors, Ages, and Metallicities}
\label{sec:colorage}

The mean $(V-I)$ color of dE GS is $(V-I) \sim 0.9$, similar to the
colors of Galactic halo GCs and to the GCs in the ``blue peak'' in
giant elliptical galaxies \cite{lotz04}.  However, several studies have
found that the mean color becomes slowly redder with increasing galaxy
luminosity \cite{strader04}\cite{lotz04}\cite{peng06}.


Recently work has been proceeding to use 8-m telescope to measure the
ages and metallicities of GCs in dEs outside the Local
Group\cite{puzia00}\cite{beasley06}\cite{conselice06}\cite{mil06}.
The metallicities fall in the range $-1.0 < [{\rm Fe/H}] < -1.5$ and
the ages are all greater than 10~Gyr.  The $[\alpha/{\rm Fe}]$ ratio
is more difficult to measure but it is important since it indicates
whether the clusters formed after a significant starburst or after a
period of quiescent star formation.  Current measurements indicate
that $[\alpha/{\rm Fe}]$ is either solar or slightly enhanced.

\begin{figure}
\begin{minipage}{\textwidth}
\includegraphics[width=6cm,angle=0]{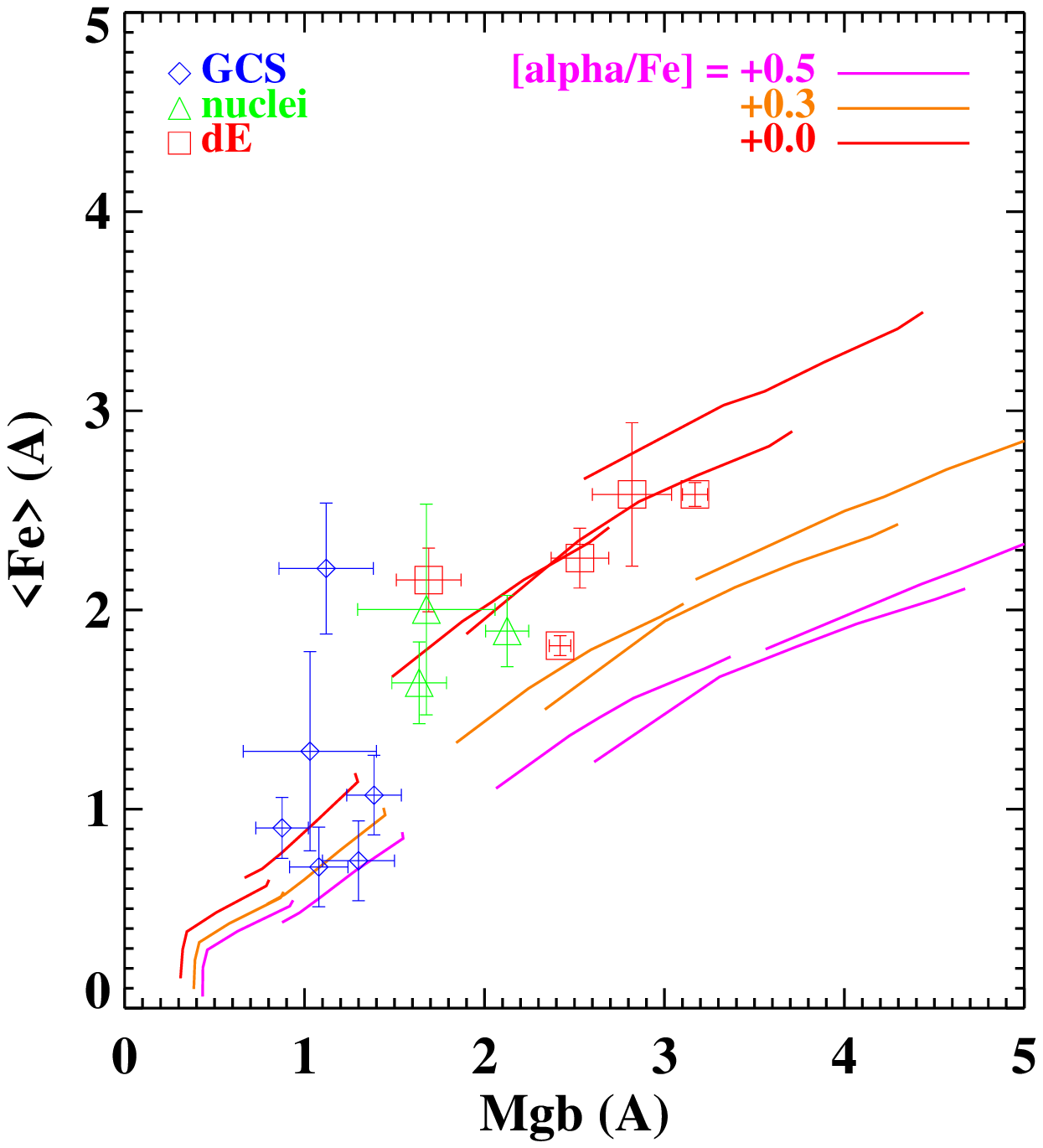} \hspace{\fill}
\includegraphics[width=6cm,angle=0]{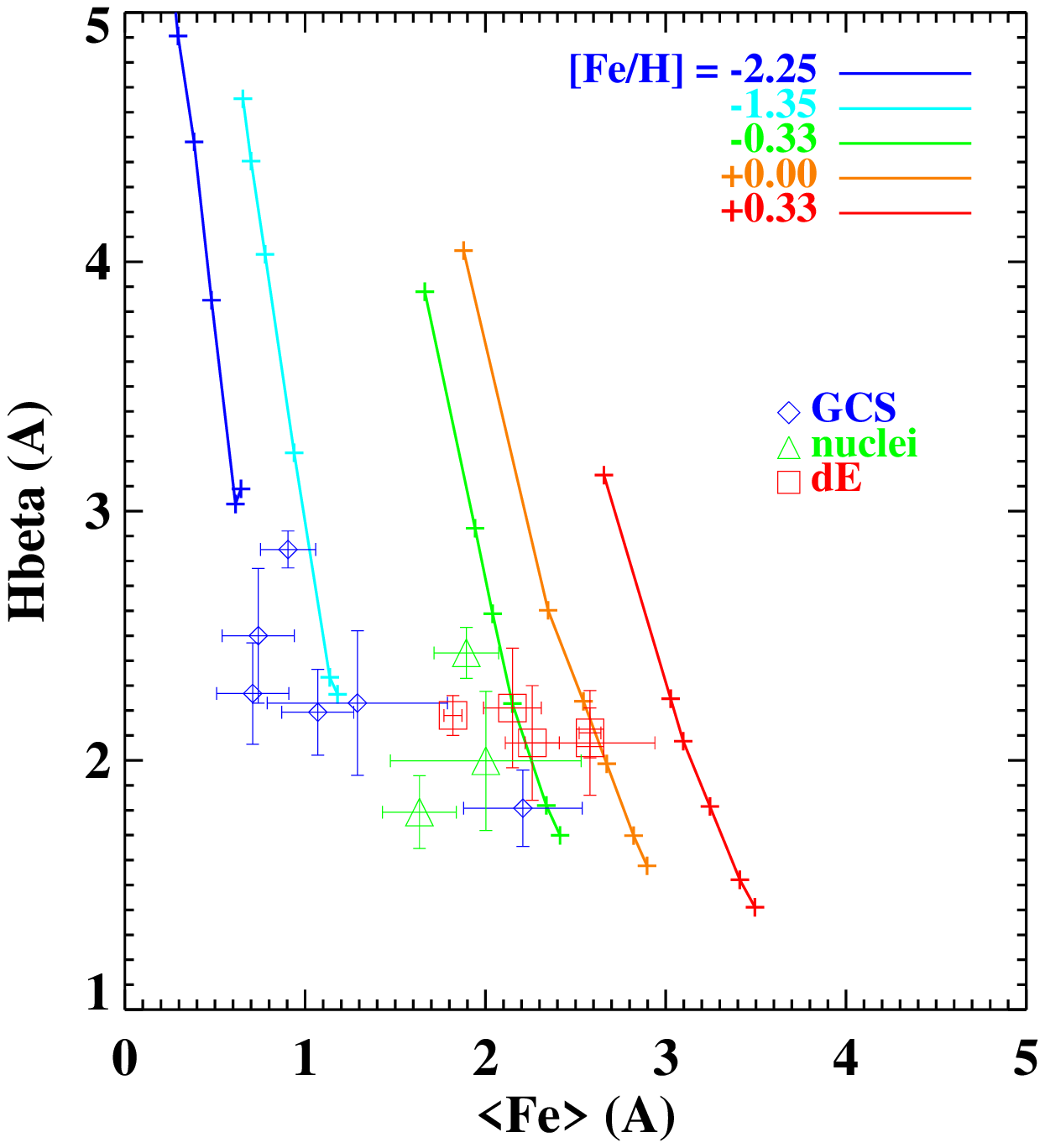} 
\end{minipage}
\caption{Line indices for GCs and nuclei from GMOS
  spectroscopy \cite{ml06} are compared with dE background
  light \cite{geha03} and stellar evolutionary models\cite{tmb03}. The
  left plot shows <Fe> versus Mgb with models for different
  $[\alpha/{\rm Fe}]$.  The right plot uses models with $[\alpha/{\rm
  Fe}]=0.0$ and shows that the GCs are old and metal-poor while the
  nuclei and dE are somewhat younger and more metal-rich.}
\label{fig:femgb}       
\end{figure}

Figure~\ref{fig:femgb} shows preliminary results of GMOS spectroscopy
of GCs and nuclei in three Virgo dEs and one Fornax dE\cite{mil06}.
The $[\alpha/{\rm Fe}]$ ratios are between 0.0 and 0.3.  Using solar
$[\alpha/{\rm Fe}]$ models we find that the ages are $>10$~Gyr and the
metallicities are $[{\rm Fe/H}] \sim -1.5$.  Interestingly, the bright
nuclei are more metal rich ($[{\rm Fe/H}] \sim -0.5$) and somewhat
younger than the typical GC. As found from photometry, the ages and
metallicities of the nuclei are intermediate between the properties of
the GCs and the background stellar light \cite{lotz04}.

\begin{figure}
\centering
\includegraphics[width=8cm,angle=0]{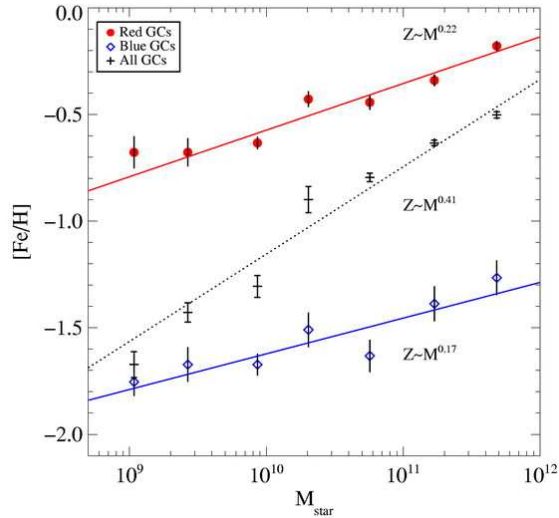}
\caption{Trends of [Fe/H] with galaxy stellar mass for red, blue, and
  all GC candidates from the ACS Virgo Survey\cite{peng06}.  The
  fraction of blue clusters increases with decreasing galaxy mass and
  the slope of the relation for the blue clusters is consistent with
  the results of \cite{strader04} and \cite{lotz04}.}
\label{fig:femass}       
\end{figure}

With the ages known the $(V-I)$ colors can be converted to
metallicities using stellar models.  The GC color--galaxy luminosity
relation then gives that $Z_{\rm MP,GC} \propto L_B^{0.2}$.  However,
the metallicity--luminosity, or mass, relation for all GCs including
the red GCs that are more common in brighter galaxies is $Z_{\rm GC}
\propto M^{0.4}$ (Figure~\ref{fig:femass})\cite{peng06}.  This is the
same as the dependence for the underlying field stars and suggests
that GCs and field stars follow a similar chemical enrichment history.

\section{Specific Frequency and GC Mass Fraction}
\label{sec:sn}

\begin{figure}
\centering
\includegraphics[width=8cm,angle=90]{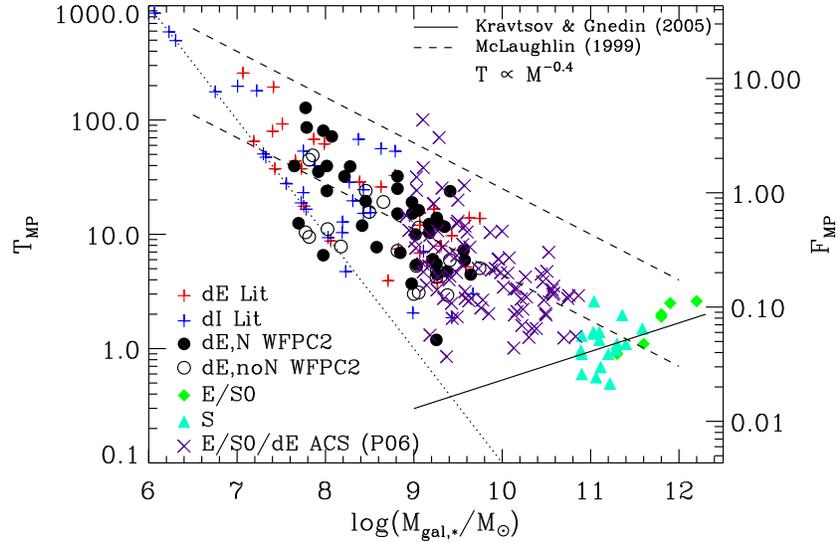}
%
%
\caption{The $T$ parameter for metal-poor (MP) GC populations
  vs. galaxy stellar mass for dE galaxies from the WFPC2 dE Snapshot
  Survey and from the literature (see \cite{ml06}).  The equivalent mass
  fraction, $F_{\rm MP}$, is shown on the right axis assuming a universal GC
  mass function.  The dotted line on the left is the line of constant
  $N_{\rm GC} = 1$.  The solid line is the prediction of $F_{\rm MP}$ with
  galaxy mass for $\log(M_{G,*}) > 10.5$ from \cite{kravtsov05}.  The
  dashed lines have a slope of $-0.4$, from the SNe-driven wind models
  of \cite{mclaughlin99}.  The lower dashed line is the prediction
  from \cite{mclaughlin99} for $M/L_V=5$ for the galaxies.  The upper
  dashed line is an approximation of the upper envelope to the
  points.}
\label{fig:tmass}       
\end{figure}

The specific globular cluster frequency is useful for
comparing globular cluster systems and it is related to the efficiency
of globular cluster formation.  In order to compare GC populations in
different type of galaxies one can calculate the $T$ parameter
\begin{equation}
T=\frac{N_{\rm GC}({\rm tot})}{M_G/10^9 M_{\odot}}
\end{equation}
which is corrected for the differences in galaxy $M/L$ \cite{za93}.
Assuming a universal GCLF (Sec.~\ref{sec:gclf}) then the GC mass
fraction is $F = 0.0433T$. Figure~\ref{fig:tmass} shows how $T$ and
$F$ for metal-poor GCs correlate with galaxy mass \cite{mil06}.  The
increase in $T$ with mass for $\log(M_{\rm gal}) > 10.5$ can be
explained by hierarchical galaxy formation models (solid
line \cite{kravtsov05}).  Below $\log(M_{\rm gal}) = 10.5$ $T$ also
increases with decreasing mass.  This can be explained by model of GC
formation that include the suppression of star formation in low-mass
halos from supernovae winds (dashed lines, \cite{mclaughlin99}).

It is also found that nucleated dEs have a mean value of $T$ about a
factor of two higher than that for non-nucleated dEs \cite{mil06}.  In
addition, dE,N galaxies are more centrally concentrated within galaxy
clusters and they have lower velocity dispersions than dE,noN
galaxies.  Therefore, the differences in $T$ values and the presence
of nuclei may be explained if dE,Ns experienced higher star formation
rates due to ``hot-mode'' gas accretion in the high density cluster
environment \cite{keres05}\cite{forbes05}.  Conversely, dE,noNs would
have formed in lower density environments where star formation is
lower due to ``cold-mode'' gas accretion.

%
%
%

\section{Conclusions}
\label{sec:conc}

Substantial progress in understanding the GCSs of dwarf galaxies has
been made in recent years due to large imaging surveys in different
environments with {\it HST}, new spectroscopic work using 8-m class
telescope, and the inclusion of globular clusters in cosmological
galaxy formation models.  



More work is still needed on photometry and spectroscopy of GCs in dIs
in order to improve the comparisons with the results on dEs.  Also,
there is much to be learned from the kinematics of GCs that could not
be discussed here.  GCs will continue to be a fundamental tool for
understanding the formation of dwarf galaxies and testing theories of
galaxy formation in general.\\

\noindent This work was supported by the Gemini Observatory, which is
operated by the Association of Universities for Research in Astronomy,
Inc., on behalf of the international Gemini partnership of Argentina,
Australia, Brazil, Canada, Chile, the United Kingdom, and the United
States of America.

%
%
%
%

%

\begin{thebibliography}{99.}

\bibitem{beasley06} M.~A.~Beasley et al.: AJ, \textbf{131}, 814 (2006)

\bibitem{baum98} H.~Baumgardt: A\&A, \textbf{330}, 480 (1998)

\bibitem{bekki06} K.~Bekki, et al: ApJL, \textbf{642}, L133 (2006)

\bibitem{conselice06} C.~J.~Conselice: ApJ, \textbf{639}, 120 (2006)


\bibitem{dicris06} M.~Di~Criscienzo, et al.:  MNRAS, \textbf{365},
  1357 (2006)

\bibitem{durrell96} P.~R.~Durrell, et al: AJ, \textbf{112}, 972 (1996)

\bibitem{fz01} S.~M.~Fall \& Q.~Zhang: ApJ, \textbf{561}, 751 (2001)

\bibitem{forbes05} D.~A.~Forbes: ApJL, \textbf{635}, L137 (2005)

\bibitem{geha03} M~Geha, et al: AJ, \textbf{126}, 1794 (2003)

\bibitem{harris86}  W.~E.~Harris: AJ, \textbf{91}, 822 (1986)


\bibitem{hg98} X.~Hernandez \& G.~Gilmore: MNRAS, \textbf{297}, 517 (1998)

\bibitem{jordan06} A.~Jord{\'a}n, et al: in these proceedings (2006)

\bibitem{keres05} D.~Kere{\v s}, et al: MNRAS, \textbf{363}, 2 (2005) 

\bibitem{kravtsov05} A.~V.~Kravtsov \& O.~Y.~Gnedin: ApJ,
  \textbf{623}, 650 (2005)

\bibitem{larsen01} S.~S.~Larsen, et al: AJ, \textbf{121}, 2974 (2001)

\bibitem{lotz01} J.~M.~Lotz, et al: ApJ, \textbf{552}, 572 (2001)

\bibitem{lotz04}  J.~M.~Lotz, B.~W.~Miller, \& H.~C.~Ferguson: ApJ,
  \textbf{613}, 262 (2004) 

\bibitem{mclaughlin99} D.~E.~McLaughlin: AJ, \textbf{117}, 2398 (1999) 

\bibitem{mil06} B.~W.~Miller, et al: in preparation (2006)

\bibitem{ml06} B.~W.~Miller \& J.~M.~Lotz: in preparation (2006)

\bibitem{minniti96} D.~Minniti, et al: A\&A, \textbf{312}, 49 (1996)

\bibitem{ohlin00} K.~S.~Oh \& D.~N.~C.~Lin: ApJ, \textbf{543}, 620 (2000)

\bibitem{peng06} E.~W.~Peng, et al: ApJ, \textbf{639}, 95 (2006)  

\bibitem{puzia00} T.~H.~Puzia, T.~H., et al: AJ, \textbf{120}, 777
(2000)

\bibitem{seth04} A.~Seth, et al: AJ, \textbf{127}, 798 (2004) 

\bibitem{sharina05} M.~E.~Sharina, et al: A\&A, \textbf{442}, 85 (2005) 

\bibitem{strader04} J.~Strader, et al: AJ, \textbf{127}, 3431 (2004)  

\bibitem{tmb03} D.~Thomas, et al:  MNRAS, \textbf{339}, 897 (2003)

\bibitem{treu05} T.~Treu, et al: ApJ, 622, L5 (2005)

\bibitem{vdb06} S.~van~den~Bergh: AJ, \textbf{131}, 304 (2006) 

\bibitem{vesp01} E.~Vesperini: MNRAS, \textbf{322}, 247 (2001)

\bibitem{za93} S.~E.~Zepf \& K.~M.~Ashman: MNRAS, \textbf{264}, 611
 (1993)

\end{thebibliography}
%



\printindex
\end{document}